# GROWTH ROUTE TOWARD III-V MULTISPECTRAL SOLAR CELLS ON SILICON

C. Renard[1], N. Cherkashin[3], A. Jaffré[2], T. Molière[1,2], L. Vincent[1], A. Michel[3], J. Alvarez[2], J. P. Connolly[4], J.P. Kleider[2], D. Mencaraglia[2], D. Bouchier[1]

[1]IEF, CNRS-UMR 8622, Bat 220, Univ Paris-Sud, 91405 Orsay, France
[2]LGEP, UMR 8507 CNRS, Supélec, Universités Paris VI et XI, 11 rue Joliot Curie, 91192 Gif-sur-Yvette
[3]CEMES-CNRS, Université de Toulouse, 29 rue J. Marvig, Toulouse, 31055, France
[4]Universidad Politécnica de Valencia, NTC, B 8F, 2°. Camino de Vera s/n. 46022, Valencia, Spain

ABSTRACT: To date, high efficiency multijunction solar cells have been developed on Ge or GaAs substrates for space applications, and terrestrial applications are hampered by high fabrication costs. In order to reduce this cost, we propose a breakthrough technique of III-V compound heteroepitaxy on Si substrates without generation of defects critical to PV applications. With this technique we expect to achieve perfect integration of heterogeneous $Ga_{1-x}In_xAs$ micro-crystals on Si substrates. In this paper, we show that this is the case for x=0. GaAs crystals were grown by Epitaxial Lateral Overgrowth on Si (100) wafers covered with a thin $SiO_2$ nanostructured layer. The cristallographic structure of these crystals is analysed by MEB and TEM imaging. Micro-Raman and Micro-Photomuminescence spectra of GaAs crystals grown with different conditions are compared with those of a reference GaAs wafer in order to have more insight on eventual local strains and their cristallinity. This work aims at developing building blocks to further develop a GaAs/Si tandem demonstrator with a potential conversion efficiency of 29.6% under AM1.5G spectrum without concentration, as inferred from our realistic modeling. This paper shows that Epitaxial Lateral Overgrowth has a very interesting potential to develop multijunction solar cells on silicon approaching the today 30.3% world record of a GaInP/GaAs tandem cell under the same illumination conditions, but on a costlier substrate than silicon.

## 1 INTRODUCTION

The growth of III-V materials on silicon substrates has been a holy grail for opto-electronics applications for many years. For photovoltaic (PV) applications in particular, this offers a very attractive alternative to III-V growth on relatively costly Ge substrates. In addition, the more affordable Si substrates are available in the form of much wider wafers (8 or 12 inches) such as those used in the microelectronics industry including PV.

One of the first fabrication routes evaluated to grow III-V on Si multijunction solar cells was the lift-off technique, which reached efficiencies of the order of 20% [1]. A more successful two-terminal approach was developed by Geisz and co-workers [2] using a graded buffer approach, but remains limited by unavoidable residual bulk defect densities.

However, the main problem when using a silicon wafer for the heteroepitaxial growth of III-V compounds is that the only one that is close to lattice match is GaP which exhibits an indirect band gap. It must be mentioned that this last drawback can be overcome by incorporating a small amount of nitrogen, the lattice match with Si being preserved with arsenic. Unfortunately, no lattice-matched III-V compounds with a gap lower than silicon exist. This precludes the full development of a long term strategy for multispectral solar cells on silicon if the route of lattice matched III-V compounds is followed, mainly due to this absence of low-gap materials.

The route that we propose is to explore and develop innovative building blocks that will permit the heteroepitaxy of mismatched III-V compounds without any substantial mechanical stress detrimental to PV applications. Therefore, the study consists in designing and optimizing the electrical junction between two relaxed lattice mismatched materials with zero or negligible strain relaxation defects, namely $Ga_{1-x}In_xAs$ and Si, in order to optimize a $GaAs/Si/Ga_{1-x}In_xAs$ multispectral solar cell demonstrator achieving a very high efficiency (close to 33%) under the global AM1.5G spectrum without concentration [3].

However, for $GaIn_{1-x}As$ monolithic epitaxy on silicon three major problems remain unresolved, namely the high density of threading dislocations in $Ga_{1-x}In_xAs$ layers grown directly on Si, the formation of anti-phase domains (APDs) due to the presence of a polar/non-polar interface and the difference in thermal expansion coefficient between the two materials [4]. In order to solve these problems, various dislocation reduction techniques have been proposed during the last years, such as thermal cycles annealing and strained-layers supperlattices as dislocation filters [4]. Nevertheless, the most promising is the epitaxial lateral overgrowth (ELO) of GaAs on Si patterned substrates with dielectric films. For this technique significant improvements have been reported for many years [5-8]. The use of nano-patterned substrates is expected to effectively remove all dislocations originating from the substrate interface. Besides, if this ELO technique is coupled with the use of a sufficiently small nucleation area size, it is expected to enable the relaxation of the mismatched material without emission of misfit dislocations [9]. The other advantage of starting from small nucleation areas is to avoid the formation of steps that contribute to the creation of APDs.

Prior know-how has been developed at IEF for $Ga_{1-x}In_xAs$ and Ge epitaxial lateral overgrowth (ELO process) on ultra-thin Si oxide by Chemical Beam Epitaxy (CBE) [10,11,12]. In this paper we demonstrate that this technique allows us to obtain perfect integration of heterogeneous GaAs islands on the micron scale size on a Si substrate. We also show that the very thin $SiO_2$ layer (0.6nm) is not detrimental to the electrical connection between the GaAs islands and the Si substrate. Further work is in progress for the integration of $Ga_{1-x}In_xAs$ with x different of 0 but here we only focus on the integration of GaAs on Si which according to the realistic modelling [3] we have performed can yield a near record efficiency tandem solar cell with a potential conversion efficiency of 29.6% under AM1.5G spectrum without concentration.

2 EXPERIMENTAL DETAILS

Epitaxial growth of GaAs micro-crystals was carried out on 4 in. p-Type Si(0 0 1) substrates in a CBE system with a base pressure of $2.10^{-8}$ Pa. Trimethylgallium (TMGa) and tertiarybutylarsine (TBAs) were used as gas sources. The Si(001) substrates have been cleaned following a modified Shiraki method. An oxide layer is formed at the last step in a solution of HCl : H2O2 : H2O (3:1:1) [13]. After this chemical cleaning step, the substrates were slowly in situ annealed up to 650°C, the pressure being maintained below $3.10^{-6}$ Pa. Prior to GaAs epitaxy, nucleation sites must be created.

Two methods were investigated to create nanoseeds, the first method, called silane ($SiH_4$) opening, consists in exposing the oxidized substrate to a silane partial pressure of 0.66 Pa at a temperature of 650°C during 4 min [14]. The second method consists in heating the oxidized substrate up to 750°C during 5 min [15] without silane exposure. These methods have the advantage of easily providing random seeds lower than 50nm in width. After this step, the GaAs epitaxy was initiated with the well-known two-step procedure [4]. The first step consists in introducing solely TBAs (during 2 min) under 430°C. In the second step TMGa was also introduced and the growth temperature was increased up to 550°C (with thermal opening only) or 575°C (with silane opening or thermal opening), respectively. After this step, GaAs epitaxial lateral overgrowth was then continued for 40 min. Then, after the GaAs islands growth at 550°C, in situ post-annealing at 600°C was performed on 1 of the 2 samples.

Post-deposition imaging of GaAs crystal morphology was obtained by using scanning electron microscopy (SEM) and transmission electron microscopy (TEM). Diffraction contrast TEM images were recorded using a Jeol 2010 TEM operating at 200 kV, from lamellas prepared by focused ion beam (FIB). The crystalline quality of microscale GaAs islands has also been analysed by micro-Raman and micro-photoluminescence (μ-PL) confocal microscopy at room temperature. The electrical junction between GaAs and Si through thin $SiO_2$ layer was investigated by Electron Beam-Induced Current (EBIC) analysis

3 RESULTS AND DISCUSSION

3.1 Growth results
Figure 1(a) and 1(b) show SEM images of microscale GaAs crystals obtained by ELO on $SiO_2$ from (001) Si nano-areas at 575°C. Figure 1(a) was taken perpendicular to the (100) plane of the Si substrate and figure 1(b) was taken with a tilt around the [110] axis. All GaAs crystals present identical shapes and have 10 {110} facets. They also present a twofold symmetry and are oriented in equal parts along the [1-10] and [110] directions according to the nominal Si wafer orientation. No change on the external form of the GaAs microcrystal was observed following the nucleation seed creation methods used.

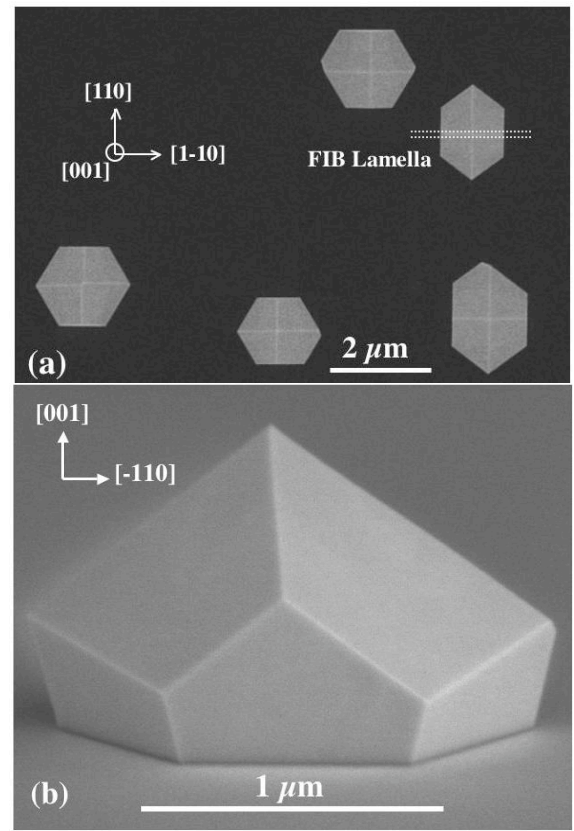

Figure 1: (a) Plan view of a scanning electron microscopy (SEM) image of microscale GaAs crystallites grown on a Si(001) surface through thin $SiO_2$ layer at 575°C. (b) tilted view around the [110] axis of a microscale GaAs crystallite.

In order to accurately study the inner structure of GaAs micro-crystals, FIB lamellas, for cross-sectional TEM observations, were prepared parallel to the two facets perpendicular to the [001] direction and was extracted approximately from the middle of the islands shown on figure 1(a). This TEM studiy has been previously published elsewhere [10].
From the TEM analysis, we can deduce that the direct GaAs/Si interface has a 55 nm width, and assume that the Si opening in the $SiO_2$ layer initially had the same width. We emphasize here that no misfit dislocation can be observed in the seed area, whereas its width is much higher than that predicted by Luryi and Suhir's theory [9]. Complementary TEM experiments have also been performed in order to determine the eventual presence of antiphase domains and no APDs are obtained. We can then confirm that nanoscale opening can avoid the APD creation, which means that the area of opening is small enough to eliminate the formation of step within it.

3.2 Micro-Raman measurements
Room temperature micro-Raman using 532 nm Nd-

YAG laser has been used to quantitatively study the crystal quality and the in-plane strain uniformity of the microscale GaAs crystals grown at 550°C and 575°C. Raman spectroscopy provides us with a simple technique to obtain information about local strain in materials, as built-in strain leads to a characteristic shift of the phonon frequency [16]. The spot size for the µ-Raman was 1µm in diameter, i.e thinner than the size of individual crystals. Consequently, each recorded µ-Raman spectrum comes from only one isolated microscale GaAs crystal.

In Figure 2, the Raman spectra of a GaAs (001) substrate and of micro crystals epitaxially grown at 550°C and 575°C are compared and found to be very different. The spectrum of the GaAs substrate used as a reference (green curve) shows mainly the LO phonon mode with a small TO contribution while spectra of micro-crystals of GaAs show solely the TO phonon mode.

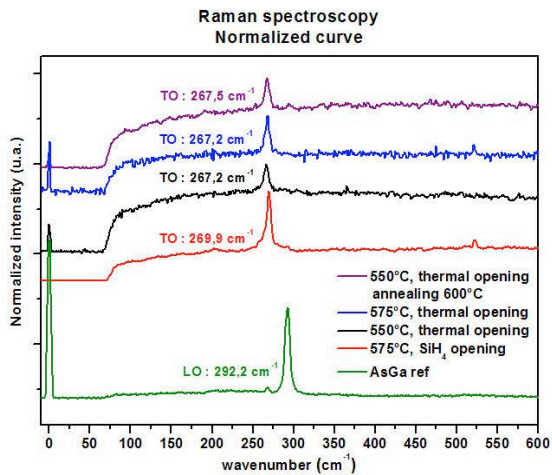

Figure 2: µ-Raman spectra measured at room temperature of GaAs (001) substrate (green curve) and of microscale GaAs crystals grown at 575°C and at 550°C.

This effect can be simply explained by the fact that the scattered intensities depend on the crystal symmetry. So, in the zinc-blende GaAs structure the TO and LO mode are split at k = 0 due to the polar nature of the crystal. As a consequence, the E1(TO) mode is allowed in backscattering from the (110) and (111) surfaces while the A1(LO) mode is allowed in backscattering from (100) and (111) surfaces, respectively [17]. As we have previously seen, the top facets of micro-crystals are mainly of (110) type. This means that the Raman spectrum of the GaAs micro-crystals can only exhibit the E1(TO) mode as the A1(LO) mode is forbidden in backscattering geometry for surfaces of the family {110}. The comparison of peak position and full width at half maximum (FWHM) measured in our µ-Raman spectra with a reference µ-Raman spectrum from a (0-11) oriented GaAs substrate should be more relevant. The data of interest can be found in a published spectrum [17] where the E1(TO) phonon mode can be measured at 267.2 cm-1. In our case, for the different microscale GaAs crystals grown at 550°C and 575°C, the E1(TO) phonon mode was found varying from 267.2 cm$^{-1}$ to 269.9 cm$^{-1}$. So, the E1(TO) phonon peak positions of microscale GaAs crystals do not reveal a significant shift with respect to the (0-11) GaAs reference. This confirms that the grown GaAs microscale crystals are pure GaAs without strain resulting from the large lattice mismatch with Si, and means that micro-crystals are fully relaxed.

3.3 Micro-Photoluminescence results

The excitation laser spot size for the Micro-Photoluminescence (µ-PL) was 5 µm in diameter, that is just slightly larger than the size of individual crystals. Consequently, the µ-PL spectrum comes from a unique microscale GaAs crystal. The excitation laser wavelength was 532 nm. The corresponding PL peak position was found varying from 1.385 eV to 1.41 eV (Figure 3). These values are comparable to the peak position at 1.426 eV, also reported in Figure 3, found for a GaAs (001) substrate with a p-type doping level of 1-5 x 10$^{17}$ cm$^{-3}$, used as reference, in the same experimental conditions.

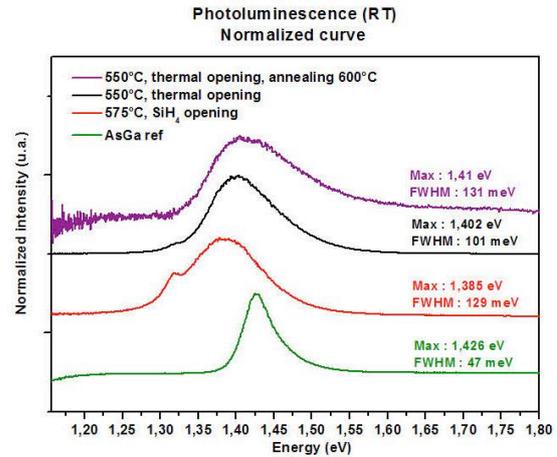

Figure 3: Room temperature micro-photoluminescence (µ-PL) spectra of different GaAs microscale islands and of a reference p-type GaAs substrate (1-5 x 10$^{17}$ cm-3)

The comparison of the peak position, of the full width at half maximum (FWHM) and of the level intensity of PL spectra with the PL of the reference GaAs substrate gives us interesting results. The intensities found from the microscale GaAs islands are comparable with that from the GaAs reference, while their FWHMs are two to three times larger (101 to 131 nm vs 47 nm). On the one hand, the relatively high intensity of the PL signal from the microscale GaAs islands clearly indicates a good crystalline quality and a low density of active defects. On the other hand, the larger FWHM suggests a doping much higher than that of the reference sample.

The comparison of the peak positions show us a redshift for all the GaAs microcrystals grown by ELO. This redshift can be explained by the fact that the observed facets are different between GaAs µ-crystals and GaAs substrate used as reference, i.e [110] or [100] respectively.
This redshift between GaAs spectra of [100] and [110] facets was previously observed by Pavesi *et al* [18]. Besides, we can clearly observe a more pronounced redshift for the GaAs µ-crystal grown at 575°C, with nucleation sites obtained with SiH$_4$ opening. This more important redshift can be attributed to a high Si doping level, as also been shown by Pavesi *et al* [18]. This significant Si doping can be easily attributed to the opening site creation with Silane. This is due to the fact that a non-negligible density of Silane decomposition products can remain at the surface of the oxide layer. Consequently, as the GaAs layer grows along the SiO2 layer by ELO process, Si atoms may be incorporated in the growing crystal and act as a dopant. Further investigations, in particular at low temperatures, to resolve the possible different contributions to the PL peak, are required to clarify this point.

TEM observations have shown that the grown GaAs microscale islands were free of APD and misfit dislocations. µ-Raman and µ-PL analysis have confirmed the good crystalline quality of GaAs microcrystals but it is questionable if the thin $SiO_2$ layers permits the electrical junction between GaAs and Si materials. This point can be evaluated via EBIC analysis.

3.4 EBIC characterization

EBIC was investigated in order to check the electrical junction between GaAs and Si materials through the thin $SiO_2$ layers. This technique employs an electron beam to induce a current within the sample which may be used as a signal for generating images that depict the electrical behaviour of the sample. With proper electrical contacts the movement of the holes and electrons generated by the SEM's electron beam can be collected, amplified, and analyzed, and be displayed as variations of contrast in an EBIC image. Two contacts were performed on the sample, a top one (on the GaAs µ-crystal) and a lower one (on the back side of the Si wafer), with nano-manipulators. In our case the interface between GaAs and Si leads to an internal electric field, and the electron–hole pairs were separated by drift due to this one.

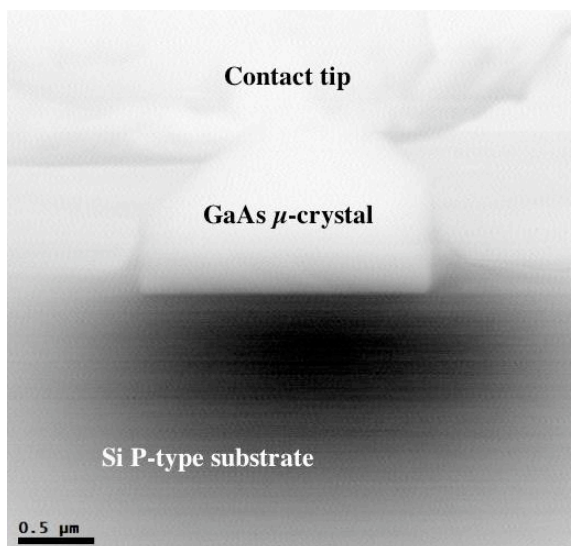

Figure 4 : EBIC map superimposed with an SEM image of the corresponding region.

This is shown in Figure 4 displaying an EBIC map superimposed with the SEM structural image of a GaAs microcrystal grown at 575°C from silane opening. The EBIC map reveals a contrast below the GaAs µ-crystal, with a maximum intensity close to its center. This indicates that the current crosses mainly through the germination hole, and to a lesser extent through the oxide area underneath the gaAs island.
Further more quantitative electrical characterizations of the GaAs/Si heterojunction are now in progress.

4 CONCLUSION

In summary, we have shown that epitaxial lateral overgrowth (ELO) of GaAs onto a thin $SiO_2$ layer from nanoscale nucleation (001) Si seeds results in the formation of high quality GaAs microscale islands, with neither misfit dislocations nor antiphase domains. µ-raman analysis confirm the good quality of GaAs crystals and from µ-PL measurements we have observed an effect of the nucleation site creation on the doping level of the GaAs µ-crystal. Finally EBIC imaging indicates that the current mainly crosses through the nucleation site, and to a lesser extent through the oxide underneath the GaAs island.
So, compared with other typical hetero-epitaxial methods of III-V growth on Si, this ELO method is therefore very promising from the point of view of its potential application for the III-V monolithic integration on Si and, as a next step, for the developpment of a GaAs/Si tandem demonstrator.

ACKNOWLEDGMENTS

This work was supported by the ANR (project MULTISOLSI n° 000901). The authors also thank the "Centrale de Technologie Universitaire" MINERVE and RENATECH for technological backup